\documentclass[runningheads]{llncs}

\def\isijcar{1}
\def\isijcar{0}

\def\appA{\if\isijcar1%
	Appendix~A of the extended version of this paper~\cite{arxiv_extended}%
	\else
	Appendix~\ref{appendix:tools_finite}%
	\fi}

\def\appB{\if\isijcar1%
	Appendix~B of the extended version of this paper~\cite{arxiv_extended}%
	\else
	Appendix~\ref{appendix:discrete}%
	\fi}

\usepackage[utf8]{inputenc}
\usepackage[english]{babel}
\usepackage[OT1]{fontenc}
\usepackage{hyperref}
\usepackage{graphicx}
\usepackage{amsfonts,amsmath}
\usepackage{cleveref}
\usepackage{scalerel}

\usepackage[colorinlistoftodos]{todonotes}
\newcommand{\dremdlw}[2][]{} 
\newcommand{\dremdk}[2][]{}

\crefname{lemmaAux}{Lemma}{Lemmas}
\crefname{theoremAux}{Theorem}{Theorems}
\crefname{definitionAux}{Definition}{Definitions}
\crefname{factAux}{Fact}{Facts}
\crefname{corollaryAux}{Corollary}{Corollarys}

\usepackage[countsame]{coqtheorem}
\setBaseUrl{{http://www.ps.uni-saarland.de/extras/fol-trakh/website/Undecidability.TRAKHTENBROT.}}

\usepackage{multicol}
\usepackage{xspace}
\usepackage{amsmath}
\usepackage{amssymb}
\usepackage{stmaryrd}
\usepackage{inferance}

\newcommand{\clap}[1]{\hbox to 0pt{\hss{#1}\hss}}

\newcommand{\MBB}[1]{\ensuremath{\mathbb{#1}}\xspace}
\newcommand{\MCL}[1]{\ensuremath{\mathcal{#1}}\xspace}
\newcommand{\MSF}[1]{\ensuremath{\mathsf{#1}}\xspace}

\newcommand{\Prop}{\MBB{P}}
\newcommand{\Type}{\MBB{T}}

\newcommand{\cdef}{\mathbin{:=}}
\newcommand{\bnfdef}{\mathbin{::=}}

\newcommand{\toot}{\mathrel\leftrightarrow}

\DeclareMathAlphabet{\mymathbb}{U}{bbold}{m}{n}
\newcommand{\Void}{\mymathbb{0}}
\newcommand{\Unit}{\mymathbb{1}}
\newcommand{\Nat}{\MBB{N}}
\newcommand{\Bool}{\MBB{B}}

\newcommand{\SigType}[2]{\Sigma{#1}.\,{#2}}
\renewcommand{\SigType}[2]{\{{#1}\mid{#2}\}}
\newcommand{\Fin}[1]{\MBB{F}_{#1}}

\newcommand{\Term}{\MSF{Term}}
\newcommand{\Formula}{\MSF{Form}}

\newcommand{\dbot}{\dot\bot}
\newcommand{\dand}{\dot\land}
\newcommand{\dor}{\dot\lor}
\newcommand{\dto}{\dot\to}
\newcommand{\dtoot}{\dot\toot}
\newcommand{\dforall}{\dot\forall}
\newcommand{\dexists}{\dot\exists}
\newcommand{\din}{\mathrel{\dot\in}}

\DeclareMathOperator*{\dbigvee}{\scalerel*{\bigvee}{\sum}}

\newcommand{\arity}[1]{\mathalpha{|{#1}|}}

\newcommand{\unit}{\mathtt{*}}
\newcommand{\btrue}{\mathsf{tt}}
\newcommand{\bfalse}{\mathsf{ff}}

\newcommand{\natS}{\mathsf{S}}
\newcommand{\some}[1]{\ulcorner#1\urcorner}
\newcommand{\none}{\emptyset}

\newcommand{\Opt}{\MBB{O}}
\newcommand{\List}{\MBB{L}}

\newcommand{\capp}{\mathbin{+\hspace{-6pt}+}} 
\newcommand{\map}{\,}
\newcommand{\inl}{\in} 
\newcommand{\incl}{\subseteq} 
\newcommand{\clen}[1]{\mathalpha{|{#1}|}}
\newcommand{\cnil}{\mathalpha{[\,]}}
\newcommand{\ccons}{\mathbin{::}}

\newcommand{\subst}[3]{{#1}[{#2}/{#3}]}

\newcommand{\tapp}{\mathbin{+\hspace{-8pt}+\hspace{-8pt}+}}

\newcommand{\red}{\mathrel\preceq}
\newcommand{\BPCP}{\MSF{BPCP}}
\newcommand{\SBPCP}{\Sigma_{\BPCP}}
\newcommand{\deriv}{\triangleright}

\newcommand{\funcssymb}{\MCL{F}}
\newcommand{\predssymb}{\MCL{P}}

\newcommand{\funcs}[1]{\funcssymb_{#1}}
\newcommand{\preds}[1]{\predssymb_{#1}}

\newcommand{\Funcs}{\funcs\Sigma}
\newcommand{\Preds}{\preds\Sigma}
\newcommand{\FV}{\mathsf{FV}}

\renewcommand{\phi}{\varphi}
\newcommand{\binop}{\,\dot\square\,}
\newcommand{\binopm}{\Box}
\newcommand{\quant}{\dot\nabla}
\newcommand{\quantm}{\nabla}

\newcommand{\MM}{\MCL{M}}
\newcommand{\BB}{\MCL{B}}

\newcommand{\SAT}{\MSF{SAT}}
\newcommand{\FSAT}{\MSF{FSAT}}
\newcommand{\FSATEQ}{\MSF{FSATEQ}}

\title{Trakhtenbrot's Theorem in Coq}
\subtitle{A Constructive Approach to Finite Model Theory}
\titlerunning{Trakhtenbrot's Theorem in Coq}
\author{Dominik Kirst$^1$ \href{https://orcid.org/0000-0003-4126-6975}{\protect\includegraphics[scale=.25]{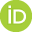}}, Dominique Larchey-Wendling$^2$
\href{https://orcid.org/0000-0001-9860-7203}{\protect\includegraphics[scale=.25]{orcid.png}}}
\authorrunning{Dominik Kirst, Dominique Larchey-Wendling}
\institute{$^1$Saarland University, Saarland Informatics Campus, Saarbrücken, Germany\\
	$^2$Universit\'e de Lorraine, CNRS, LORIA, Vandœuvre-l\`es-Nancy, France\\
	\path|kirst@ps.uni-saarland.de|\hspace{1cm}    	\path|dominique.larchey-wendling@loria.fr|}

\newenvironment{proofqed}{\begin{proof}}{\qed\end{proof}}

\begin{document}
\maketitle

\begin{abstract}
We study finite first-order satisfiability (FSAT)
in the constructive setting of dependent type theory.
Employing synthetic accounts of enumerability and decidability, 
we give a full classification of FSAT depending on
the first-order signature of non-logical symbols.
On the one hand, 
our development focuses on Trakhtenbrot's theorem, stating
that FSAT is undecidable as soon as the signature contains an at least binary 
relation symbol. Our proof proceeds by a many-one reduction chain 
starting from the Post correspondence problem.
On the other hand, we establish the decidability of FSAT for monadic 
first-order logic, i.e.\ where the signature only 
contains at most unary function and relation symbols, as well as the enumerability 
of FSAT for arbitrary 
enumerable signatures.
All our results are mechanised in the framework of a growing Coq library of synthetic undecidability proofs.
\end{abstract}

\section{Introduction}

In the wake of the seminal discoveries concerning 
the undecidability of first-order logic by Turing and Church in the 1930s, 
a broad line of work has been pursued to characterise the border between decidable 
and undecidable fragments of the original decision problem.
These fragments can be grouped either by syntactic restrictions 
controlling the allowed 
function and relation symbols
or the quantifier prefix, or by semantic restrictions 
on the admitted models (see~\cite{borger1997classical} for
a comprehensive description).

Concerning signature restrictions, already predating the undecidability results, Löwenheim had shown in 1915 that monadic first-order logic, admitting only signatures with at most unary symbols, is decidable~\cite{Lowenheim1915}.
Therefore, the successive negative results usually presuppose non-trivial signatures containing an at least binary symbol.

Turning to semantic restrictions, Trakhtenbrot proved in 1950 that, if only admitting finite models, the satisfiability problem over non-trivial signatures is still undecidable~\cite{trakhtenbrot50}.
Moreover, the situation is somewhat dual to the unrestricted case, since finite satisfiability (FSAT) is still enumerable while, in the unrestricted case, validity is enumerable.
As a consequence, finite validity cannot be characterised by a complete finitary deduction system and, resting on finite model theory, various natural problems in database theory are undecidable.

Conventionally, Trakhtenbrot's theorem is proved by (many-one) reduction from the halting problem for Turing machines (see e.g.~\cite{borger1997classical,Libkin:2010:EFM:1965351}).
An encoding of a given Turing machine $M$ can be given as a formula $\phi_M$ such that the models of $\phi_M$ correspond to the runs of $M$.
Specifically, the finite models of $\phi_M$ correspond to terminating runs of $M$ and so a decision procedure for finite satisfiability of $\phi_M$ would be enough to decide whether $M$ terminates or not.

Although this proof strategy is in principle explainable on paper, already the formal definition of Turing machines, not to mention their encoding in first-order logic, is not ideal for mechanisation in a proof assistant.
So for our Coq mechanisation of Trakhtenbrot's theorem, we follow a different strategy by starting from the Post correspondence problem (PCP), a simple matching problem on strings.
Similar to the conventional proof, we proceed by encoding every instance $R$ of PCP as a formula $\phi_R$ such that $R$ admits a solution iff $\phi_R$ has a finite model.
Employing the framework of synthetic undecidability~\cite{ForsterCPP,library_coqpl}, 
the computability of $\phi_R$ from $R$ is guaranteed since all functions definable in constructive type theory are computable without reference to a concrete model of computation.

Both the conventional proof relying on Turing machines and our elaboration starting from PCP actually produce formulas in a custom signature well-suited for the encoding of the seed decision problems.
The sharper version of Trakhtenbrot's theorem, stating that a signature with at least one binary relation (or one binary function and one unary relation) is enough to turn FSAT undecidable, is in fact left as an exercise in e.g.~Libkin's book~\cite{Libkin:2010:EFM:1965351}.
However, at least in a constructive setting, this generalisation is non-trivial and led us to mechanising a chain of signature transformations eliminating and compressing function and relation symbols step by step.

Complementing the undecidability result, we further formalise that FSAT is enumerable for enumerable signatures and decidable for monadic signatures.
Again, both of these standard results come with their subtleties when explored in a constructive approach of finite model theory.

In summary, the main contributions of this paper are threefold:
\begin{itemize}
	\item
	we provide an axiom-free Coq mechanisation comprising a full classification of finite satisfiability with regards to the signatures allowed;\footnote{Downloadable from \url{http://www.ps.uni-saarland.de/extras/fol-trakh/} and systematically hyperlinked with the definitions and theorems in this PDF.}
	\item
	we present a streamlined proof strategy for Trakhtenbrot's theorem well-suited for mechanisation and simple to explain informally, basing on PCP;
	\item
	we give a constructive account of signature transformations and the treatment of interpreted equality typically neglected in a classical development.
\end{itemize}

The rest of the paper is structured as follows.
We first describe the type-theoretical framework for undecidability proofs and the representation of first-order logic in Section~\ref{sec:prelims}.
We then outline our variant of Trakhtenbrot's theorem for a custom signature in Section~\ref{sec:trakh_prelim}.
This is followed by a development of enough constructive finite model theory (Section~\ref{sec:finmod}) to conclude some decidability results (Section~\ref{sec:decidability}) as well as the final classification (Section~\ref{sec:trakh_full}).
We end with a brief discussion of the Coq development and future work in Section~\ref{sec:discussion}.

\section{First-Order Satisfiability in Constructive Type Theory}
\label{sec:prelims}

In order to make this paper accessible to readers unfamiliar with constructive type theory, we outline the required features of Coq's underlying type theory, the synthetic treatment of computability available in constructive mathematics, some properties of finite types, as well as our representation of first-order logic.

\subsection{Basics of Constructive Type Theory}

We work in the framework of a constructive type theory such as the one implemented in Coq, providing a predicative hierarchy of type universes $\Type$ above a single impredicative universe $\Prop$ of propositions.
On type level, we have the unit type $\Unit$ with a single element $\unit:\Unit$, 
the void type $\Void$, function spaces $X\to Y$, products $X\times Y$, sums $X+Y$, dependent products $\forall x:X.\,F\,x$, and dependent sums $\SigType{x:X}{F\,x}$.
On propositional level, these types are denoted using the usual logical notation ($\top$, $\bot$, $\to$, $\land$, $\lor$, $\forall$, and $\exists$).

We employ the basic inductive types
of Booleans ($\Bool\bnfdef \btrue\mid\bfalse$), of
Peano natural numbers ($n:\Nat\bnfdef 0\mid\natS\,n$), 
the option type ($\Opt\,X\bnfdef \some{x}\mid\none$),
and lists ($l:\List\,X\bnfdef\cnil\mid x\ccons l$).
We write $\clen l$ for the \emph{length} of a list, $l\capp m$ for the \emph{concatenation}
of $l$ and $m$, $x\inl l$ for \emph{membership}, 
and simply $f\map [x_1;\ldots;x_n]\cdef [f\,x_1;\ldots;f\,x_n]$ for the 
\emph{map} function.
We denote by $X^n$ the type of vectors of length $n:\Nat$ 
and by $\Fin n$ the finite types understood as indices
$\{0,\ldots,n-1\}$.
The definitions/notations for lists are shared with vectors $\vec v:X^n$.
Moreover, when $i:\Fin n$ and $x:X$, we denote by $\vec v_i$ the $i$-th component of $\vec v$
and by $\subst{\vec v}xi$ the vector $\vec v$ with $i$-th component updated to value $x$.
%

\subsection{Synthetic (Un-)decidability}

We review the main ingredients of our synthetic approach to 
decidability and undecidability~\cite{forster2018verification,ForsterCPP,forster2019certified,library_coqpl,Larchey-WendlingForster:2019:H10_in_Coq,SpiesForster:2019:UndecidabilityHOU}, based on the computability of all 
functions definable in constructive type theory\rlap.%
\footnote{A result shown and applied for many variants of constructive type theory 
and which Coq designers are committed to maintain as Coq evolves.}\enspace
We first introduce standard notions of computability theory without referring to a formal model of computation, e.g.\ Turing machines.

\begin{definition}
\label{def:decidable}
	A \emph{problem} or predicate $p:X\to \Prop$ is
	\begin{itemize}
		\item
		\setCoqFilename{decidable}
		\emph{\coqlink[decidable_bool_eq]{decidable}} if there is $f:X \to\Bool$ with
		$\forall x.\,p\,x\toot f\,x=\btrue$.
		\item
		\setCoqFilename{enumerable}
		\emph{\coqlink[opt_enum_t]{enumerable}} if there is $f:\Nat\to\Opt\,X$ with $\forall x.\,p\,x\toot \exists n.\, f\,n=\some x$.
	\end{itemize}
	These notions generalise to predicates of higher arity.
	Moreover, a type $X$ is
	\begin{itemize}
		\item
		\setCoqFilename{enumerable}
		\emph{\coqlink[type_enum_t]{enumerable}} if there is $f:\Nat\to\Opt\,X$ with $\forall x. \exists n.\, f\,n=\some x$.
		\item
		\setCoqFilename{decidable}
		\emph{\coqlink[discrete]{discrete}} if equality on $X$ (i.e.\ $\lambda xy:X.\,x=y$) is decidable.
		\item
		a \emph{data type} if it is both enumerable and discrete.
	\end{itemize}
\end{definition}

\setCoqFilename{decidable}
Using the expressiveness of dependent types, we equivalently tend to establish the decidability 
of a predicate $p:X\to\Prop$ by giving a function \coqlink[decidable]{$\forall x:X.\,p\,x +\neg p\,x$}.
Note that it is common to mechanise decidability results in this synthetic sense (e.g.~\cite{braibant2010efficient,maksimovic2015hocore,schafer2015completeness}).
Next, decidability and enumerability transport along reductions:

\setBaseUrl{{http://www.ps.uni-saarland.de/extras/fol-trakh/website/Undecidability.}}
\setCoqFilename{Problems.Reduction}
\begin{definition}[][reduces]
	A problem $p:X\to\Prop$ \emph{(many-one) reduces} to $q:Y\to\Prop$, written $p\red q$, 
        if there is a function $f: X\to Y$ such that $p\,x\toot q\,(f\,x)$ for all $x:X$.%
        \footnote{Or equivalently, the 
        \coqlink[reduction_dependent]{dependent characterisation} $\forall x:X.\,\SigType{y:Y}{p\,x\toot q\,y}$.}
\end{definition}

\setBaseUrl{{http://www.ps.uni-saarland.de/extras/fol-trakh/website/Undecidability.TRAKHTENBROT.}}
\begin{fact}
	\label{fact:reduction}
	Assume\/ $p:X\to\Prop$, $q:Y\to\Prop$ and\/ $p\red q$:
        \setCoqFilename{decidable}%
        \coqlink[reduction_decidable]{(1)}~if\/ $q$ is decidable, then so is\/ $p$ and
        \setCoqFilename{enumerable}%
	\coqlink[reduction_opt_enum_t]{(2)}~if\/ $X$ and\/ $Y$ are data types and\/ $q$ is enumerable, 
                then so is\/ $p$.
\end{fact}


Item~(1) implies that we can justify the undecidability of a target problem by reduction from a seed problem known to be undecidable, such as the halting problem for Turing machines.
This is in fact the closest rendering of undecidability available in a synthetic setting, since the underlying type theory is consistent with the assumption that every problem is decidable\rlap.\footnote{As witnessed by classical set-theoretic models satisfying $\forall p:\Prop.\, p+\neg p$ ({cf.}~\cite{werner_sets_1997}).}\enspace
Nevertheless, we believe that in the intended effective interpretation for synthetic computability, a typical seed problem is indeed undecidable and so are the problems reached 
by verified reductions\rlap.\footnote{This synthetic treatment of undecidability is discussed in more detail in~\cite{ForsterCPP} and \cite{library_coqpl}.}\enspace
More specifically, since the usual seed problems are not co-enumerable, (2) implies that the reached problems are not co-enumerable either.

\smallskip

Given its simple inductive characterisation involving only basic types of lists and Booleans, 
the (binary) Post correspondence problem (\BPCP) is a well-suited seed problem for compact encoding into first-order logic.

\setBaseUrl{{http://www.ps.uni-saarland.de/extras/fol-trakh/website/Undecidability.TRAKHTENBROT.}}
\setCoqFilename{bpcp}
\begin{definition}
	Given a list $R: \List(\List\,\Bool\times \List\,\Bool)$ of pairs $s/t$ of Boolean strings\rlap,\footnote{Notice that
        the list $R$ is viewed as a (finite) set of pairs $s/t\inl R$ (hence ignoring the order or duplicates), 
        while $s$ and $t$, which are also lists, are viewed a strings (hence repetitions and ordering matter for $s$ and
        $t$).}   
        we define \emph{derivability} of a pair $s/t$ from $R$ (denoted by \coqlink[pcp_hand]{$R\deriv s/t$})  
        and \emph{solvability} (denoted by \coqlink[BPCP_problem]{$\BPCP\,R$}) by the following rules:
	\[\infer1{s/t \inl R}{R\deriv s/t}                               \quad\qquad
	\infer2{s/t \inl R}{R\deriv u/v}{R\deriv (s\capp u)/(t\capp v)}  \quad\qquad
	\infer1{R\deriv s/s}{\BPCP\,R}
        \]
\end{definition}

\begin{fact}
  \label{fact:BPCP_undec}
        Given a list\/ $R:\List(\List\,\Bool\times \List\,\Bool)$, the derivability predicate $\lambda s\,t.R\deriv s/t$ 
        is \coqlink[bpcp_hand_dec]{decidable}. However, the halting problem for Turing machines reduces to\/ \BPCP.
\end{fact}

\setCoqFilename{red_utils}
\begin{proofqed}
        We give of proof of the decidability of $R\deriv s/t$ by induction on $\clen s+\clen t$.
        We also provide a \coqlink[BPCP_BPCP_problem_eq]{trivial proof} of the equivalence of two definitions of \BPCP.
        See~\cite{forster2018verification,forster2019certified} for details on the reduction from the halting problem to \BPCP. 
\end{proofqed}

It might at first appear surprising that derivability $\lambda s\,t.R\deriv s/t$ is decidable while \BPCP\ is reducible from the halting
problem (and hence undecidable). This simply illustrates that undecidability is caused by the unbounded existential quantifier in the equivalence 
$\BPCP\,R\toot \exists s.\,  R\deriv s/s$.

\subsection{Finiteness}



\setBaseUrl{{http://www.ps.uni-saarland.de/extras/fol-trakh/website/Undecidability.Shared.Libs.DLW.Utils.}}
\setCoqFilename{fin_base}
\begin{definition}
	A \emph{type $X$ is \coqlink[finite_t]{finite}} if there is a list $l_X$ with $x\inl l_X$ for all $x:X$ and
	a \emph{predicate $p:X\to\Prop$ is \coqlink[fin_t]{finite}} if there is a list $l_p $ with $\forall x.\, p\,x\toot x\inl l_p $.
\end{definition}

Note that in constructive settings there are various alternative characterisations of finiteness%
\footnote{And these alternative characterisations are not necessarily constructively equivalent.}
(bijection with $\Fin n$ for some $n$; negated infinitude for some definition of infiniteness; etc.)
and we opted for the above since it is easy to work with while transparently capturing the expected meaning. 
One can distinguish \emph{strong} finiteness in $\Type$ (i.e.\ $\SigType{l_X:\List\,X}{\forall x.\, x\inl l_X}$) 
from \emph{weak} finiteness in $\Prop$ (i.e.\ $\exists l_X:\List\,X.\,\forall x.\, x\inl l_X$), the list $l_X$ being required 
computable in the strong case.

\smallskip

We present three important tools for manipulating finite types: the 
 \emph{finite pigeon hole principle} (PHP) here established without assuming discreteness, 
the \emph{well-foundedness} of strict orders over finite types, and \emph{quotients} over strongly decidable 
equivalences that map onto $\Fin n$.
The proofs are given in \appA.

For the finite PHP, the typical classical proof requires the discreteness 
of $X$ to design transpositions/permutations. Here we avoid discreteness 
completely, the existence of a duplicate being established without 
actually computing one.

\setCoqFilename{php}
\begin{theorem}[Finite PHP][PHP_rel]
\label{thm:finite_php}
Let\/ $R:X\to Y\to\Prop$ be a binary relation 
and\/ $l:\List\,X$ and $m:\List\,Y$ be two lists where\/
$m$ is shorter than\/ $l$ $(\clen m < \clen l)$. 
If\/ $R$ is total from\/ $l$ to\/ $m$ 
$(\forall x.\, x\inl l \to \exists y.\, y\inl m \land R\,x\,y)$ then
the values at two distinct positions in\/ $l$ are related to the same\/ $y$ in\/ $m$,
i.e.\/ there exist\/ $x_1,x_2\inl l$ and $y\inl m$ such that\/ $l$ has shape\/ 
$l=\cdots\capp x_1\ccons \cdots\capp x_2\ccons \cdots$ and\/ $R\,x_1\,y$ and\/ $R\,x_2\,y$.
\end{theorem}

Using the PHP, one can constructively show that,
for a strict order over a finite type $X$,
any descending chain has length bounded 
by the size of $X$.\footnote{i.e.\ the length of the enumerating list of $X$.}

\setBaseUrl{{http://www.ps.uni-saarland.de/extras/fol-trakh/website/Undecidability.}}
\setCoqFilename{Shared.Libs.DLW.Wf.wf_finite}
\begin{fact}[][wf_strict_order_finite]
	\label{fact:wf}
	Every strict order on a finite type is well-founded.
\end{fact}

Coq's type theory does not provide quotients in general (see e.g.~\cite{cohen13}) but one can build 
computable quotients in certain conditions, here for a decidable equivalence
relation of which representatives of equivalence classes are listable.

\setCoqFilename{Shared.Libs.DLW.Utils.fin_quotient}
\begin{theorem}[Finite decidable quotient][decidable_EQUIV_fin_quotient]
\label{thm:fin_quotient}
Let\/ ${\sim}:X\to X\to\Prop$ be a decidable equivalence with
$\SigType{l_r:\List\,X}{\forall x\exists y.\,y\inl l_r\land x\sim y}$, i.e.\/
finitely many equivalence classes.\footnote{Hence $l_r$ denotes a list of \emph{representatives} of equivalence classes.}\enspace
Then one can compute the quotient\/ $X/{\sim}$ onto\/ $\Fin n$ for some $n$, i.e.\/
$n:\Nat$,\/ $c:X\to\Fin n$ and\/ $r:\Fin n\to X$ s.t.\/
$\forall p.\,c\,(r\,p) = p$ and\/ $\forall x \,y.\,x\sim y\toot c\,x = c\, y$.
\end{theorem}

Using Theorem~\ref{thm:fin_quotient} with identity over $X$ as equivalence,
we get bijections between finite, discrete types and the type family 
$(\Fin n)_{n:\Nat}$.\footnote{For a given $X$, the value $n$ (usually called cardinal) is unique
by the PHP.}

\setCoqFilename{Shared.Libs.DLW.Utils.fin_bij}
\begin{corollary}[][finite_t_discrete_bij_t_pos]
\label{coro:fin_type}
If\/ $X$ is a finite and discrete type then one can compute\/~$n:\Nat$
and a bijection from\/ $X$ to\/ $\Fin n$.
\end{corollary}

\subsection{Representing First-Order Logic}

\setBaseUrl{{http://www.ps.uni-saarland.de/extras/fol-trakh/website/Undecidability.TRAKHTENBROT.}}

We briefly outline our representation of the syntax and semantics of first-order logic in constructive type theory ({cf.}~\cite{ForsterEtAl:2019:Completeness}).
Concerning the \emph{syntax}, we describe terms and formulas as dependent inductive types over a
\setCoqFilename{fo_sig}\coqlink[fo_signature]{signature} $\Sigma=(\Funcs; \Preds)$ of function
symbols $f:\Funcs$ and relation symbols $P:\Preds$ with arities $\arity f$ and $\arity P$, 
using binary  connectives ${\binop}\in\{{\dto},{\dand},{\dor}\}$ and
quantifiers ${\quant}\in\{{\dforall},{\dexists}\}$:
%
%
		$$\begin{array}{r@{\,:\,}l@{~\bnfdef~}l@{\qquad}l}
                  t &
                  \setCoqFilename{fo_terms}
                  \coqlink[fo_term]{\Term_\Sigma} & x \mid f\,\vec{t}  &  (x:\Nat,~ f:\Funcs,~  \vec t:\Term^{\arity f}_\Sigma\,) \\
		\phi,\psi & 
                  \setCoqFilename{fo_logic}
                  \coqlink[fol_form]{\Formula_\Sigma} & \dbot\mid P\,\vec{t}
                                   \mid \phi \binop \psi
                                   \mid \quant\phi 
                   & (P:\Preds,~ \vec t:\Term^{\arity P}_\Sigma\,)
                 \end{array}$$
	Negation is defined as the abbreviation $\dot{\neg}\phi\cdef \phi\,\dto\,\dbot$.

\setCoqFilename{fo_logic}

In the chosen de Bruijn representation~\cite{de_bruijn_lambda_1972}, a bound variable is encoded as the number of 
quantifiers shadowing its binder, e.g. $\forall x.\,\exists y.\, P\,x\,u\to P\,y\,v$ may be represented 
by $\dforall\,\dexists\,P\,1\,4\,\dto\, P\,0\,5$.
The variables $2 = 4-2$ and $3 = 5-2$ in this example are the \emph{free} variables, and variables that do not occur freely are called \emph{fresh},
e.g.\ $0$ and $1$ are fresh.
For the sake of legibility, we write concrete formulas with named binders and defer de Bruijn representations to the Coq development.
For a formula $\phi$ over a signature $\Sigma$, we define 
the list $\FV(\phi):\List\,\Nat$ of \coqlink[fol_vars]{free variables},
the list $\funcs\phi : \List\,\Funcs$ of \coqlink[fol_syms]{function symbols} 
and the list $\preds\phi : \List\,\Preds$ of \coqlink[fol_rels]{relation symbols} 
that actually occur in $\phi$, all by recursion on $\phi$.

%

Turning to \emph{semantics}, we employ the standard (Tarski-style) model-theoretic semantics, evaluating terms in a given domain and embedding the logical connectives into the constructive meta-logic ({cf.}~\cite{veldman_models}):

\setBaseUrl{{http://www.ps.uni-saarland.de/extras/fol-trakh/website/Undecidability.TRAKHTENBROT.}}
\setCoqFilename{fo_sig}
\begin{definition}
	A \emph{\coqlink[fo_model]{model}} $\MM$ over a domain $D:\Type$ is described by a pair of functions
        $\forall f.\,D^{|f|}\to D$ and $\forall P.\,D^{|P|}\to \Prop$ denoted by $f^\MM$ and $P^\MM$.
        \setCoqFilename{fo_terms}
	Given a \emph{variable assignment} $\rho:\Nat\to D$, we recursively extend it to a \emph{\coqlink[fo_term_sem]{term evaluation}} $\hat\rho:\Term\to D$ with $\hat\rho \,x\cdef \rho\,x$ and $\hat\rho\,(f\,\vec v)\cdef f^\MM\,(\hat{\rho}\map\vec v)$,
        \setCoqFilename{fo_logic}
	and to the \emph{\coqlink[fol_sem]{satisfaction}} relation $\MM\vDash_\rho \phi$ by
	\begin{align*}
	\MM\vDash_\rho \dot{\bot}&~\cdef ~\bot&
	\MM\vDash_\rho \phi\binop\psi &~\cdef ~\MM\vDash_\rho\phi~\binopm~ \MM\vDash_\rho\psi\\
	\MM\vDash_\rho P\,\vec t\,&~\cdef ~P^\MM\,(\hat{\rho}\map\vec t\,)&
	\MM\vDash_\rho\quant\phi&~\cdef ~\quantm a:D.\,\MM\vDash_{a\cdot\rho} \phi
	\end{align*}
 	where each logical connective $\binop$/$\quant$ is mapped to its meta-level counterpart 
        $\binopm$/$\quantm$ and 
        \setCoqFilename{notations}%
        where we denote by $a\cdot\rho$ the \coqlink[de_bruijn_ext]{de Bruijn extension} of $\rho$ by $a$, defined by
        $(a\cdot\rho)\,0\cdef a$ and $(a\cdot\rho)\,(1+x)\cdef\rho\,x$.\footnote{The notation $a\cdot\rho$ 
        illustrates that $a$ is pushed ahead of the sequence $\rho_0,\rho_1,\ldots$}
\end{definition}

A \emph{$\Sigma$-model} is thus a dependent triple $(D,\MM,\rho)$ composed of a domain $D$, a
model $\MM$ for $\Sigma$ over $D$ and an assignment $\rho:\Nat\to D$. It is \emph{finite} if $D$ is finite, and
\emph{decidable} if $P^\MM:D^{|P|}\to \Prop$ is decidable for all $P:\Preds$.

\setCoqFilename{fo_logic}
\begin{fact}[][fol_sem_dec]
\label{fact:satisfaction_dec}
Satisfaction\/ 
$\lambda \phi.\,\MM\vDash_\rho \phi$ is decidable for finite, decidable\/ $\Sigma$-models.
\end{fact}

\begin{proofqed}
By induction on $\phi$; finite quantification preserves decidability.
\end{proofqed}

In this paper, we are mostly concerned with \emph{finite satisfiability} of formulas.
However, since some of the compound reductions hold for more general or more specific notions, 
we introduce the following variants:

\setCoqFilename{fo_sat}
\begin{definition}[Satisfiability]
	For a formula $\phi$ over a signature $\Sigma$, we write
	\begin{itemize}
		\item $\SAT(\Sigma)\,\phi$ if there is a\/ $\Sigma$-model\/ $(D,\MM,\rho)$ such that $\MM\vDash_\rho \phi$;
		\item \coqlink[fo_form_fin_dec_SAT]{$\FSAT(\Sigma)\,\phi$} if additionally $D$ is finite and $\MM$ is decidable;
		\item \coqlink[fo_form_fin_dec_eq_SAT]{$\FSATEQ(\Sigma;\equiv)\,\phi$} if the signature contains a distinguished binary relation symbol $\equiv$ 
                interpreted as equality, i.e.\/ $x\equiv^\MM y\toot x=y$ for all $x,y:D$.
	\end{itemize}
\end{definition}

%
Notice that in a classical treatment of finite model theory, models are supposed to be given \emph{in extension},
i.e.\ understood as tables providing computational access to functions and relations values.
To enable this view in our constructive setting, we restrict to decidable relations in the definition of $\FSAT$, and from now on,
\emph{finite satisfiability is always meant to encompass a decidable model}.
One could further require the domain $D$ to be discrete to conform more closely with the classical view;
discreteness is in fact enforced by \FSATEQ. 
However, we refrain from this requirement and instead show in Section~\ref{sec:discrete_models} that $\FSAT$
and $\FSAT$ over discrete models are constructively equivalent.
\dremdk{My idea was to hide FSAT' a bit so the reader don't have to track too many notions. That's why I'd prefer to state theorems like 17 with FSAT and use the equivalence from 15 wlog. in the proof.}

\section{Trakhtenbrot's Theorem for a Custom Signature}
\label{sec:trakh_prelim}

In this section, we show that \BPCP reduces to $\FSATEQ(\SBPCP;{\equiv})$ for the special purpose signature $\SBPCP\cdef (\{\star^0, e^0, f_\btrue^1, f_\bfalse^1\}; \{P^2, {\prec^2},{\equiv^2}\})$.
To this end, we fix an instance $R:\List\,(\List\,\Bool\times \List\,\Bool)$ of $\BPCP$ (to be understood 
as a finite set of pairs of Boolean strings) and 
we construct a formula $\phi_R$ such that $\phi_R$ is finitely satisfiable if and only if $R$ has a solution.

Informally, we axiomatise a family $\BB_n$ of  models over the domain of Boolean strings of length bounded by $n$ and let $\phi_R$ express that $R$ has a solution in $\BB_n$.
The axioms express enough equations and inversions of the constructions included in the definition of $\BPCP$ such that a solution for $R$ can be recovered.

Formally, the symbols in $\SBPCP$ are used as follows:
the functions $f_b$ and the constant $e$ represent $b\ccons(\cdot)$ and $\cnil$ 
for the encoding of strings $s$ as terms $\overline s$:
$$\overline{\cnil}\tapp \tau\cdef  \tau\hspace{3em}
\overline{b\ccons s}\tapp \tau\cdef f_b\,(\overline s\tapp \tau)\hspace{3em}
\overline s\cdef  \overline s\tapp e$$
The constant $\star$ represents an undefined value for strings too long to be encoded in the finite model $\BB_n$.
The relation $P$ represents derivability from $R$ (denoted $R\deriv\cdot/\cdot$ here) while $\prec$ and $\equiv$ represent strict suffixes and equality, respectively.

Expected properties of the intended interpretation can be captured formally as first-order formulas.
First, we ensure that $P$ is proper (only subject to defined values) and that $\prec$ is a strict order (irreflexive and transitive):
$$
\begin{array}{c@{~\cdef~}l@{\qquad}l}
\phi_P     & \dforall x y.\, P\,x\,y ~\dto~ x\not\equiv \star ~\dand~ y \not\equiv\star & \text{($P$  proper)} \\
\phi_\prec & (\dforall x.\, x\not\prec x)~\dand~ (\dforall x y z.\, x\prec y~\dto~ y\prec z~\dto~ x\prec z) & \text{($\prec$ strict order)}\\
\end{array}$$
Next, the image of $f_b$ is forced disjoint from $e$ and injective as long as $\star$ is not reached.
We also ensure that the images of $f_\btrue$ and $f_\bfalse$ intersect only at $\star$:
$$\phi_f~\cdef~\left(\begin{array}{@{\,}l@{\,}}
  f_\btrue\,\star \equiv \star ~\dand~ f_\bfalse\,\star \equiv \star\\
  \dforall x.\, f_\btrue\,x\not\equiv e\\
  \dforall x.\, f_\bfalse\,x\not\equiv e \\
\end{array}\right)
\,\dand\,
\left(\begin{array}{@{\,}l@{\,}}
  \dforall x y.\,f_\btrue\,x\not\equiv \star~\dto~ f_\btrue\,x\equiv f_\btrue\,y~\dto~ x\equiv y \\
  \dforall x y.\,f_\bfalse\,x\not\equiv \star~\dto~ f_\bfalse\,x\equiv f_\bfalse\,y~\dto~ x\equiv y \\
  \dforall x y.\,f_\btrue \,x\equiv f_\bfalse \,y~\dto~ f_\btrue \,x\equiv \star ~\dand~ f_\bfalse \,y\equiv \star\\
\end{array}\right)$$
Furthermore, we enforce that $P$ simulates $R\deriv\cdot/\cdot$, encoding its inversion principle
%
	$$\phi_\deriv\cdef \dforall x y.\, P\,x\,y ~\dto\,\dbigvee^{\boldsymbol .}_{\clap{\scriptsize$ s/t\!\inl\! R$}}\, 
        \dor\left\{\begin{array}{@{\,}l} x\equiv  \overline s ~\dand~ y \equiv \overline t\\
	                                \dexists u v.\,P\,u\,v~\dand~ x\equiv\overline s\tapp u~\dand~ y\equiv\overline t\tapp v~\dand~ u/v\prec x/y\\
                   \end{array}\right.$$
%
where $u/v\prec x/y$ denotes $(u\prec x~\dot\land~ v\equiv y)\dot\lor (v\prec y~\dot\land~ u\equiv x)\dot\lor (u\prec x ~\dot\land~ v\prec y)$.
Finally, $\phi_R$ is the conjunction of all axioms plus the existence of a solution:
$$\phi_R \cdef  \phi_P ~\dot\land~ \phi_\prec ~\dot\land~ \phi_f ~\dot\land~ \phi_\deriv ~\dot\land~ \dot\exists x.\, P\,x\,x.$$

\setCoqFilename{red_undec}
\begin{theorem}[][BPCP_FIN_DEC_EQ_SAT]
	\label{thm:BPCP_FSATEQ}
	$\BPCP\red \FSATEQ(\SBPCP;{\equiv})$.
\end{theorem}

\begin{proofqed}
The reduction $\lambda R.\,\phi_R$ is proved correct by
Lemmas~\ref{lemma:BCPC_FSATEQ} and~\ref{lemma:FSATEQ_BPCP}.
\end{proofqed}

\setCoqFilename{BPCP_SigBPCP}
\begin{lemma}[][Sig_bpcp_encode_sound]
        \label{lemma:BCPC_FSATEQ}
	$\BPCP\,R\to \FSATEQ(\SBPCP;{\equiv})\,\phi_R$.
\end{lemma}

\begin{proofqed}
	Assume $R\deriv s/s$ holds for a string $s$ with $|s|= n$.
	We show that the model $\BB_n$ over Boolean strings bounded by $n$ satisfies $\phi_R$.
	To be more precise, we choose $D_n\cdef \Opt \SigType{s:\List\Bool}{\clen s\le n}$
	as domain, i.e.\ values in $D_n$ are either an (overflow) value $\none$ or a (defined) dependent pair
        $\some{(s,H_s)}$ where $H_s:\clen s\le n$. We interpret the function and relation symbols of the chosen signature by
	\begin{align*}
	e^{\BB_n}&\cdef \cnil & f_b^{\BB_n}\,\none&\cdef  \none & P^{\BB_n}\,s\,t&\cdef R\deriv s/t\\
	\star^{\BB_n}&\cdef \none & f_b^{\BB_n}\,s&\cdef \textnormal{if $\clen s< n$ then $b\ccons s$ else $\none$} &
	s\prec^{\BB_n} t&\cdef s\not = t\land \exists u.\,u\capp s=t
	\end{align*}
	where we left out some explicit constructors and the excluded edge cases of the relations for 
        better readability.
	As required, $\BB_n$ interprets $\equiv$ by equality $=_{D_n}$.
	
	Considering the desired properties of $\BB_n$, first note that $D_n$ can be shown finite by induction 
        on $n$. This however crucially relies on the proof irrelevance of the $\lambda x.\,x\le n$ predicate\rlap.%
        \footnote{i.e.\ that for every $x:\Nat$ and $H,H':x\le n$ we have $H=H'$. In general, it is
        not always possible to establish finiteness of $\SigType x{P\,x}$ if $P$ is not proof irrelevant.}\enspace
	The atoms $s\prec^{\BB_n} t$ and $s\equiv^{\BB_n} t$ are decidable by straightforward computations 
        on Boolean strings. Decidability of $P^{\BB_n}s\,t$ (i.e.\ $R\deriv s/t$) was established in Fact~\ref{fact:BPCP_undec}.
	Finally, since $\phi_R$ is a closed formula, any variable assignment 
        $\rho$ can be chosen to establish that $\BB_n$ satisfies $\phi_R$, for instance $\rho\cdef\lambda x.\none$.
	Then showing $\BB_n\vDash_\rho\phi_R$ consists of verifying simple properties of the chosen functions and relations, 
        with mostly straightforward proofs.
\end{proofqed}

\begin{lemma}[][Sig_bpcp_encode_complete]
        \label{lemma:FSATEQ_BPCP}
	$\FSATEQ(\SBPCP;{\equiv})\,\phi_R\to \BPCP\,R$.
\end{lemma}

\begin{proofqed}
	Suppose that $\MM\vDash_\rho\phi_R$ holds for some finite $\SBPCP$-model $(D,\MM,\rho)$ interpreting $\equiv$ as equality 
        and providing operations $f^\MM_b$, $e^\MM$, $\star^\MM$, $P^\MM$ and $\prec^\MM$.
	Again, the concrete assignment $\rho$ is irrelevant and $\MM\vDash_\rho\phi_R$ ensures that the functions/relations 
        behave as specified and that $P^\MM\,x\,x$ holds for some $x:D$.
	
	Instead of trying to show that $\MM$ is isomorphic to some $\BB_n$, we directly reconstruct a solution for $R$,
        i.e.\ we find some $s$ with $R\deriv s/s$ from the assumption that $\MM\vDash_\rho\phi_R$ holds.
	To this end, we first observe that the relation $u/v\prec^\MM x/y$ as defined above is a strict order and
        thus well-founded as an instance of Fact~\ref{fact:wf}.
	
	Now we can show that for all $x/y$ with $P^\MM\,x\,y$ there are strings $s$ and $t$ with $x=\overline s$, $y=\overline t$ 
        and $R\deriv s/t$, by induction on the pair $x/y$ using the well-foundedness of  $\prec^\MM$.
	So let us assume $P^\MM\,x\,y$. Since $\MM$ satisfies $\phi_\deriv$ there are two cases:
	
	\begin{itemize}
		\item
		there is $s/t\in R$ such that $x=\overline s$ and $y =\overline t$.
		The claim follows by $R\deriv s/t$;
		\item
		there are $u,v:D$ with $P^\MM\,u\,v$ and $s/t\in R$ such that $x=\overline s\tapp u$,  $y=\overline t\tapp v$, and $u/v\prec^\MM x/y$.
		The latter makes the inductive hypothesis applicable for $P^\MM\,u\,v$, hence yielding $R\deriv s'/t'$ for some 
                strings $s'$ and $t'$ corresponding to the encodings $u$ and $v$.
		This is enough to conclude $x=\overline{s\capp s'}$, $y=\overline{t\capp t'}$ and $R\deriv(s\capp s')/(t\capp t')$ as wished.
	\end{itemize}
	
	\noindent
	Applying this fact to the assumed match $P^\MM\,x\,x$ yields a solution $R\deriv s/s$.
\end{proofqed}

\section{Constructive Finite Model Theory}
\label{sec:finmod}

\newcommand{\foequiv}[1]{\mathrel{\dot =_{#1}}}
\newcommand{\foindist}{\foequiv{}}
\newcommand{\genrel}{\mathcal R}
\newcommand{\lf}{l_\funcssymb}
\newcommand{\lp}{l_\predssymb}

\newcommand{\Fb}{\mathrm F}
\newcommand{\Ff}{\Fb_\funcssymb}
\newcommand{\Fp}{\Fb_\predssymb}

\newcommand{\bisim}{\mathrel{\equiv_\Fb}}
\newcommand{\dequiv}{\mathrel{\dot\equiv}}

Combined with Fact~\ref{fact:BPCP_undec}, Theorem~\ref{thm:BPCP_FSATEQ} entails the undecidability (and non-co-enumerability)
of \FSATEQ over a custom (both finite and discrete) signature $\SBPCP$. 
By a series of signature reductions, we generalise these results to any signature containing 
an at least binary relation symbol. In particular, we explain how to reduce $\FSAT(\Sigma)$ 
to $\FSAT(\Void; \{\in^2\})$ for any discrete signature $\Sigma$, hence including $\SBPCP$.
We also provide a reduction from $\FSAT(\Void; \{\in^2\})$ to $\FSAT(\{f^n\};\{P^1\})$ 
for $n\geq 2$, which entails the undecidability of $\FSAT$ for signatures with 
one unary relation and an at least binary function.
But first, let us show that $\FSAT$ is unaltered when further assuming 
discreteness of the domain. 

\subsection{Removing Model Discreteness and Interpreted Equality}

\label{sec:discrete_models}

We consider the case of models over a discrete domain $D$.
Of course, in the case of $\FSATEQ(\Sigma;{\equiv})$ the requirement that
$\equiv$ is interpreted as a decidable binary relation which is equivalent to $=_D$ 
imposes the discreteness of $D$. 
But in the case of $\FSAT(\Sigma)$ nothing imposes such a restriction on $D$.
However, as we argue here, we can always quotient $D$ using 
a suitable decidable congruence,
making the quotient a discrete finite type while preserving 
first-order satisfaction.

\setCoqFilename{fo_sat}
\begin{definition}[][fo_form_fin_discr_dec_SAT]
	We write $\FSAT'(\Sigma)\,\phi$ if $\FSAT(\Sigma)\,\phi$ on a discrete model.
\end{definition}

\dremdk{I like the detailed explanation but I think this might be too lengthy for the flow of the paper. How about defining indistinguishability here and just stating that it can be obtained as a finite fixpoint maintaining decidability with the detailed construction in an appendix?}

Let us consider a fixed signature $\Sigma=(\Funcs; \Preds)$.
In addition, let us fix a finite type $D$ and a (decidable) model $\MM$
of $\Sigma$ over $D$.
We can conceive an equivalence over $D$ which is
a congruence for all the interpretations of the symbols
by $\MM$, namely \emph{first-order indistinguishability}
$x\foequiv\Sigma y~\cdef~\forall \phi\,\rho.\, \MM\vDash_{x\cdot\rho}\phi\toot \MM\vDash_{y\cdot\rho}\phi$,
i.e.\ first-order semantics in $\MM$ is not impacted when switching 
$x$ with $y$. 

The facts that $\foequiv\Sigma$ is both an equivalence and a congruence are easy to prove but, 
with this definition, there is little hope of establishing decidability of $\foequiv\Sigma$. 
The main reason for this is that the signature may contain symbols of infinitely many arities.
So we fix two lists $\lf  : \List\,\Funcs$ and $\lp  : \List\,\Preds$ of
function and relation symbols respectively and restrict the congruence
requirement to these lists.

\setCoqFilename{discrete}
\begin{definition}[Bounded first-order indistinguishability][fo_bisimilar]
We say that\/ $x$ and\/ $y$ are \emph{first-order indistinguishable up to\/ $\lf /\lp $}, 
and we write $x\foindist y$, if for any $\rho:\Nat\to D$ and any first-order formula\/ $\phi$ built
from the symbols in\/ $\lf $ and\/ $\lp $ only, we have
$\MM\vDash_{x\cdot\rho}\phi\toot \MM\vDash_{y\cdot\rho}\phi$.
\end{definition}

\begin{theorem}[][fo_bisimilar_dec_congr]
\label{thm:fo_indist}
First-order indistinguishability\/ $\foindist$ up to\/ $\lf /\lp $ is a strongly decidable
equivalence and a congruence for all the symbols in\/ $\lf /\lp $. 
\end{theorem}

\begin{proofqed}
The proof is quite involved, we only give its sketch here; 
see \appB\ for more details.
The real difficulty is to show the decidability of $\foindist$. To this end,
we characterise $\foindist$ as a bisimulation, i.e.\ we show that $\foindist$ is extensionally equivalent to Kleene's greatest fixpoint
$\Fb^\omega(\lambda uv.\top)$
of some $\omega$-continuous operator~$\Fb:(D\to D\to \Prop) \to (D\to D\to \Prop)$. We then show that
$\Fb$ preserves strong decidability. To be able to conclude, we establish that
$\Fb$ reaches its limit after $l\cdef 2^{d\times d}$ iterations where $d\cdef\mathrm{card}\,D$,
the length of a list enumerating the finite type $D$. To verify this upper bound, we
build the \emph{weak powerset}, a list of length $l$ which contains all the weakly decidable 
binary predicates of type $D\to D\to \Prop$, up to extensional equivalence. As all the iterated values
$\Fb^n(\lambda uv.\top)$ are strongly decidable, they all belong to
the weak powerset, so by Theorem~\ref{thm:finite_php},
a duplicate is to be found in the first $l+1$ steps, ensuring that
the sequence is stalled at $l$.
\end{proofqed}

We use the strongly decidable congruence $\foindist$ to quotient models onto
discrete ones (in fact $\Fin n$ for some $n$) while preserving first-order satisfaction. 

\setCoqFilename{red_utils}
\begin{theorem}[][fo_form_fin_dec_SAT_discr_equiv]
        \label{thm:quotient_FSAT}
	For every first-order signature\/ $\Sigma$ and formula\/ $\phi$ over\/ $\Sigma$, 
        we have\/ $\FSAT(\Sigma)\,\phi$ iff\/ $\FSAT'(\Sigma)\,\phi$,
        and as a consequence, both reductions\/
        $\FSAT(\Sigma)\red \FSAT'(\Sigma)$ and\/
        $\FSAT'(\Sigma)\red \FSAT(\Sigma)$ hold.
\end{theorem}

\begin{proofqed}
$\FSAT(\Sigma)\,\phi$ entails $\FSAT'(\Sigma)\,\phi$ is the non-trivial implication.
Hence we consider a finite $\Sigma$-model $(D,\MM,\rho)$ of $\phi$ and we build a new
finite $\Sigma$-model of $\phi$ which is furthermore discrete.
We collect the symbols occurring in
$\phi$ as the lists $\lf \cdef\funcs\phi$ (for functions) and $\lp \cdef\preds\phi$ (for relations).
By Theorem~\ref{thm:fo_indist}, first-order indistinguishability ${\foindist}:D\to D\to\Prop$  
up to $\funcs\phi/\preds\phi$ is  a strongly decidable equivalence over $D$ and a congruence for the semantics of the symbols occurring in $\phi$.
Using Theorem~\ref{thm:fin_quotient}, we build the quotient $D/{\foindist}$ on a $\Fin n$ 
for some $n:\Nat$. We transport the model $\MM$ along this quotient
and because $\foindist$ is a congruence for the symbols in $\phi$, its
semantics is preserved along the quotient. Hence, $\phi$ has a finite 
model over the domain $\Fin n$ which is both finite and discrete.
\end{proofqed}

\begin{theorem}[][FIN_DEC_EQ_SAT_FIN_DEC_SAT]
	\label{thm:uninterpret}
        If\/ $\equiv$ is a binary relation symbol in the
        signature\/ $\Sigma$, one has a
        reduction\/ $\FSATEQ(\Sigma;{\equiv})\red \FSAT(\Sigma)$.
\end{theorem}

\begin{proofqed}
Given a list $\lf$ (resp.\ $\lp$) of function (resp.\ relation) symbols,
we construct a formula $\psi(\lf,\lp,{\equiv})$ over the function symbols 
in $\lf $ and relation symbols in $({\equiv}\ccons \lp )$ expressing
the requirement that $\equiv$ is
an equivalence and a congruence for the symbols in $\lf /\lp $. 
Then we show that
$\lambda\phi.\, \phi~\dand~\psi(\funcs\phi, {\equiv}::\preds\phi,{\equiv})$
is a correct reduction, where $\funcs\phi$ and $\preds\phi$
list the symbols occurring in $\phi$.
%
%
\end{proofqed}

\subsection{From Discrete Signatures to Singleton Signatures}

Let us start by converting a discrete signature to a finite and discrete 
signature.

\setCoqFilename{Sig_Sig_fin}
\begin{lemma}[][Sig_discrete_to_pos]
\label{lem:signature_finite}
For any formula\/ $\phi$ over a discrete signature\/ $\Sigma$, 
one can compute a 
signature\/ $\Sigma_{n,m}=(\Fin n;\Fin m)$,
arity preserving maps $\Fin n\to \Funcs$ and $\Fin m\to \Preds$
and an \emph{equi-satisfiable} formula\/ $\psi$ over\/  $\Sigma_{n,m}$, i.e.\/ 
$\FSAT(\Sigma)\,\phi\toot\FSAT(\Sigma_{n,m})\,\psi$.
\end{lemma} 

\begin{proofqed}
We use the discreteness of $\Sigma$ and bijectively map the lists of symbols $\funcs\phi$ and $\preds\phi$ 
onto $\Fin n$ and $\Fin m$ respectively, using 
Corollary~\ref{coro:fin_type}. We structurally map $\phi$ to $\psi$  
over $\Sigma_{n,m}$ along this bijection, which preserves finite satisfiability.
\end{proofqed}


Notice that $n$ and $m$ in the signature $\Sigma_{n,m}$ depend on $\phi$, hence
the above statement cannot be presented as a reduction between (fixed) signatures.

\smallskip

We now erase all function symbols by encoding them with relation symbols.
To this end, let $\Sigma=(\Funcs; \Preds)$ be a signature, we set $\Sigma'\cdef(\Void; \{\equiv^2\}+\Funcs^{+1}+\Preds)$
where $\equiv$ is a new interpreted relation symbol of arity two and in the conversion, function symbols have arity lifted by
one, hence the $\Funcs^{+1}$ notation.

\setCoqFilename{red_undec}
\begin{lemma}[][FIN_DISCR_DEC_SAT_FIN_DEC_EQ_NOSYMS_SAT]
	\label{lem:remove_functions}
        For any finite\/\footnote{In the Coq code, we prove the theorem for finite \emph{or} discrete types of function symbols.} 
        type of function symbols\/ $\Funcs$, one has a
        reduction\/ $\FSAT'(\Funcs; \Preds)\red \FSATEQ(\Void; \{\equiv^2\}+\Funcs^{+1}+\Preds;{\equiv^2})$.
\end{lemma}

\begin{proofqed}
The idea is to  recursively replace a term $t$ over $\Sigma$ by a formula which is ``equivalent'' to 
$x\equiv t$ (where $x$ is a fresh variable not occurring in $t$)
and then an atomic formula like e.g.\ $P\,[t_1;t_2]$ by  $\exists\, x_1\,x_2.\, x_1\equiv t_1\,\dand\,x_2\equiv t_2\,\dand\, P\,[x_1;x_2]$. 
We complete the encoding with a formula stating that every function symbol $f:\Funcs$ is encoded into a
total functional relation $P_f: \Funcs^{+1}$ of arity augmented by $1$.
\end{proofqed}


Next, assuming that the function symbols have already been erased, we explain how to merge the relation symbols in a signature 
$\Sigma=(\Void; \Preds)$ into a single relation symbol, provided that there is an upper bound for the arities in $\Preds$.

\begin{lemma}[][FSAT_REL_BOUNDED_ONE_REL]
\label{lem:bounded_arity}
The reduction\/ $\FSAT(\Void; \Preds)\red\FSAT\bigl(\Void; \{Q^{1+n}\}\bigr)$ holds
when\/ $\Preds$ is a finite and discrete type of relation symbols and\/
$|P|\le n$ holds for all\/ $P:\Preds$.
\end{lemma}


\begin{proofqed}
This comprises three independent reductions, see Fact~\ref{prop:three_reds} below.
\end{proofqed}

In the following, we denote
by $\Funcs^n$ (resp.\ $\Preds^n$) the same type of function (resp.\ 
relation) symbols but where the arity is uniformly converted to $n$.

\begin{fact}
  \label{prop:three_reds}
  Let $\Sigma=(\Funcs; \Preds)$ be a signature:
  \begin{enumerate}
  \coqitem[FSAT_UNIFORM] $\FSAT(\Funcs; \Preds)\red \FSAT(\Funcs; \Preds^n)$ if\/ $|P|\le n$ holds for all\/ $P:\Preds$;
  \coqitem[FSAT_ONE_REL] $\FSAT(\Void; \Preds^n)\red \FSAT(\Preds^0;\{Q^{1+n}\})$ if\/ $\Preds$ is finite;
  \coqitem[FSAT_NOCST] $\FSAT(\Funcs^0; \Preds)\red \FSAT(\Void; \Preds)$ if\/ $\Funcs$ is discrete.
  \end{enumerate}
\end{fact}

\begin{proofqed}
For the first reduction, every atomic formula of the form $P\,\vec v$ with $\clen{\vec v}=\arity P\leq n$ is converted to $P\,\vec v'$ with $\vec v'\cdef \vec v \capp [x_0;\dots;x_0]$ and $\clen{\vec v'}= n$ for an arbitrary term variable $x_0$.
The rest of the structure of formulas is unchanged.

For the second reduction, we convert every atomic formula $P\,\vec v$ with $\clen{\vec v}=n$
into $Q(P\ccons\vec v)$ where $P$ now represents a constant symbol
($Q$ is fixed). 

For the last reduction, we replace every constant symbol by a corresponding fresh variable
chosen above all the free variables of the transformed formula.
\end{proofqed}

\subsection{Compressing $n$-ary Relations to Binary Membership}

\label{sec:compressing}

Let $\Sigma_n=(\Void;\{P^n\})$ be a singleton signature where $P$ is of arity $n$.
We now show that $P$ can be compressed to a binary relation modelling 
membership via a construction using hereditarily finite 
sets~\cite{SmolkaStark:2016:Hereditarily}
(useful only when $n\ge 3$).

\begin{theorem}[][FIN_DISCR_DEC_nSAT_FIN_DEC_2SAT]
	\label{thm:compress_relations}
	$\FSAT'(\Void; \{P^n\})\red \FSAT(\Void; \{\din^2\})$.
\end{theorem}

Technically, this reduction is one of the most involved in this work, although in most
presentations of Trakhtenbrot's theorem, this is left as an ``easy exercise\rlap,'' see 
e.g.~\cite{Libkin:2010:EFM:1965351}. Maybe it is perceived so because it relies
on the encoding of tuples in set theory, which is somehow natural for 
mathematicians\rlap,\footnote{In our case we use Kuratowski's encoding.} but 
properly building the finite set model 
in constructive type theory was not that easy. 

Here we only give an overview
of the main tools.
We encode an arbitrary $n$-ary relation $R:X^n\to\Prop$ over a finite type $X$
in the theory of \emph{membership} over the signature $\Sigma_2=(\Void; \{{\din}^2\})$.
Membership is much weaker than set theory because the only required
set-theoretic axiom is \emph{extensionality}. Two sets are extensionally equal if
their members are the same, and extensionality states that two extensionally equal sets 
belong to the same sets:
\begin{equation}
\label{eq:ext}
\dforall x y.\, (\dforall z.\, z\din x~\dtoot~z\din y)~\dto~\dforall z.\, x\din z~\dto~y\din z
\end{equation}
As a consequence, no first-order formula over $\Sigma_2$ can distinguish two extensionally 
equal sets. 
Notice that the language of membership theory (and set theory) does not contain
any function symbol, hence, contrary to usual mathematical practices, there is no
other way to handle a set than via its characterising formula which makes it
a very cumbersome language to work with formally. However, this is how we have to proceed 
in the Coq development but here, we stick to meta-level ``terms'' in the prose for simplicity. 

\newcommand{\pair}[2]{({#1},{#2})}
\newcommand{\myred}{\Sigma_{n\rightsquigarrow 2}}
\newcommand{\myredr}{\myred^r}
\newcommand{\powerset}{\mathcal P}

\setCoqFilename{Sign_Sig2}

The ordered pair of two sets $p$ and $q$ is encoded as $\pair p q \cdef \{\{p\},\{p,q\}\}$
while the $n$-tuple $(t_1,\ldots,t_n)$ is encoded as $\pair{t_1}{(t_2,\ldots,t_n)}$
recursively. The reduction function which maps formulas  over $\Sigma_n$ to
formulas over $\Sigma_2$ proceeds as follows. We reserve two first-order 
variables $d$ (for the domain $D$) and $r$ (for the relation $R$). We describe 
the recursive part of the reduction \coqlink[Sign_Sig2_encoding]{$\myredr$}
$$\begin{array}{@{}r@{~\cdef~}l@{~~}r@{~\cdef~}l@{}}
  \myredr(P\,\vec v) & \text{``$\mathrm{tuple}\,\vec v\din r$''}
& \myredr(\dforall z.\,\phi) & \dforall z.\,z\din d~\dto~\myredr(\phi)\\
  \myredr(\phi\,\binop\,\psi) & \myredr(\phi)\,\binop\,\myredr(\psi)
& \myredr(\dexists z.\,\phi) & \dexists z.\,z\din d~\dand~\myredr(\phi)\\
\end{array}
$$
ignoring the de Bruijn syntax (which would imply adding $d$ and $r$ as
parameters). Notice that $d$ and $r$ should not occur freely in $\phi$.
In addition, we require that:
$$
\begin{array}{r@{~\cdef~}l@{\qquad}l}
\phi_1 & \text{$\din$ is extensional} & \text{see Equation~\eqref{eq:ext};}\\
\phi_2 & \dexists z.\, z\din d        & \text{i.e.\ $d$ is non-empty;}\\
\phi_3 & x_1\din d~\dand~\cdots~\dand~x_k\din d & \text{where $[x_1;\ldots;x_k] = \FV(\phi)$.}
\end{array}$$
This gives us the reduction function $\myred(\phi) \cdef \phi_1\,\dand\,\phi_2\,\dand\,\phi_3\,\dand\,\myredr(\phi)$.

\smallskip

The \emph{\coqlink[SAT2_SATn]{completeness}}
of the reduction $\myred$ is the easy part. Given a finite model of $\myred(\phi)$
over $\Sigma_2$, we recover a model of $\phi$ over $\Sigma_n$ by selecting as the new
domain the members of $d$ and the interpretation of $P\,\vec v$ is given by
testing whether the encoding of $\vec v$ as a $n$-tuple is a member of $r$. 

The \emph{\coqlink[SATn_SAT2]{soundness}} of the reduction $\myred$ is the formally involved part, with Theorem~\ref{thm:Sign_Sig2}
below containing the key construction. 

\setCoqFilename{reln_hfs}
\begin{theorem}[][reln_hfs]
\label{thm:Sign_Sig2}
Given a decidable\/ $n$-ary relation\/ $R:X^n\to\Prop$ over a finite, discrete
and inhabited type $X$, one can compute a finite and discrete type\/ $Y$ equipped 
with a decidable relation ${\in}:Y\to Y\to\Prop$, two distinguished
elements\/ $d,r:Y$ and a pair of maps\/ $i:X\to Y$ and\/ $s:Y\to X$ s.t.\par
\medskip
\centerline{$\displaystyle
\begin{array}{@{}ll@{\quad}ll@{}}
  \text{1.} & \text{$\in$ is extensional;}
& \text{4.} & \text{$\forall x:X.\,i\,x\in d$;} \\
  \text{2.} & \text{extensionally equal elements of\/ $Y$ are equal;}
& \text{5.} & \text{$\forall y:Y.\,y\in d\to\exists x.\, y=i\,x$;} \\
  \text{3.} & \text{all\/ $n$-tuples of members of\/ $d$ exist in\/ $Y$;}
& \text{6.} & \text{$\forall x:X.\, s(i\,x) = x$;}\\[0.3ex]
  \multicolumn{4}{@{}l}{\text{7.}~\text{$R\,\vec v$ iff\/ $i(\vec v)$ is a\/ $n$-tuple member of\/ $r$, for any\/ $\vec v:X^n$.}}\\
\end{array}$}
\end{theorem}

\begin{proofqed}
	We give a brief outline of this quite involved proof, referring to the \coqlink[reln_hfs]{Coq code} for details.
	The type $Y$ is built from the type of hereditarily finite sets based on~\cite{SmolkaStark:2016:Hereditarily},
	and when we use the word ``set'' below, it means hereditarily finite set.
	The idea is first to construct $d$ as a \emph{transitive set} of which the elements are in bijection $i/s$
	with the type $X$, hence $d$ is the cardinal of $X$ in the set-theoretic meaning.
	Then the iterated powersets $\powerset(d), \powerset^2(d),\ldots,\powerset^k(d)$ are all transitive as well and contain $d$
	both as a member and as a subset. Considering $\powerset^{2n}(d)$
	which contains all the $n$-tuples built from the members of  $d$, we
	define $r$ as the set of $n$-tuples collecting the encodings $i(\vec v)$ of vectors
	$\vec v:X^n$ such that $R\,\vec v$. We show $r\in p$ for $p$ defined as
	$p\cdef \powerset^{2n+1}(d)$. 
        Using the Boolean counterpart of $(\cdot)\in p$ for unicity of proofs,
        we then define $Y\cdef \{z\mid z\in p\}$,
	restrict membership $\in$ to $Y$ and this gives the finite type equipped 
	with all the required properties. Notice that the decidability requirement for ${\in}$
        holds constructively because we work with hereditarily finite sets, and would not hold with arbitrary sets.
\end{proofqed}

\subsection{Summary: From Discrete Signatures to the Binary Signature}

Combining all the previous results, 
we give a reduction from any discrete signature
to the binary singleton signature.

\setCoqFilename{red_undec}
\begin{theorem}[][DISCRETE_TO_BINARY]
\label{thm:discrete_to_binary}
$\FSAT(\Sigma)\red \FSAT(\Void; \{P^2\})$ holds for any discrete signature\/ $\Sigma$.
\end{theorem}

\begin{proofqed}
Let us first consider the case of $\Sigma_{n,m}=(\Fin n;\Fin m)$, a signature
over the finite and discrete types $\Fin n$ and $\Fin m$. 
Then we have a reduction $\FSAT(\Fin n;\Fin m)\red\FSAT(\Void; \{P^2\})$
by combining Theorems~\ref{thm:quotient_FSAT}, \ref{thm:uninterpret} and~\ref{thm:compress_relations} 
and Lemmas~\ref{lem:remove_functions} and~\ref{lem:bounded_arity}.

Let us denote by $f_{n,m}$ the reduction $\FSAT(\Fin n;\Fin m)\red\FSAT(\Void; \{P^2\})$.
Let us now consider a fixed discrete signature $\Sigma$. For a formula $\phi$ over $\Sigma$, 
using Lemma~\ref{lem:signature_finite}, we compute a signature $\Sigma_{n,m}$
and $\psi$ over $\Sigma_{n,m}$ s.t.\ $\FSAT(\Sigma)\,\phi\toot\FSAT(\Fin n;\Fin m)\,\psi$.
The map $\lambda\phi.f_{n,m}\,\psi$ is the required reduction. 
\end{proofqed}

\begin{lemma}[][FSAT_REL2_to_FUNnREL1]
	\label{lem:sig2_sig21}
	$\FSAT(\Void; \{P^2\})\red \FSAT(\{f^n\};\{Q^1\})$ when\/ $n\geq 2$.
\end{lemma}

\setCoqFilename{Sig2_SigSSn1}

\begin{proofqed}
We encode the binary relation $\lambda x\,y.\,P\,[x;y]$ with $\lambda x\,y.\,Q \bigl(f\,[x;y;\dots]\bigr)$, 
using the first two parameters of $f$ to encode pairing. But since we need to change the 
domain of the model, we also use a fresh variable $d$ to encode the domain as 
$\lambda x.\,Q (f\,[d;x;\dots])$ and we \coqlink[Sig2_SigSSn1_encoding]{restrict all quantifications to the domain}
similarly to the encoding $\myredr$ of Section~\ref{sec:compressing}.
\end{proofqed}

We finish the reduction chains with the weakest possible signature constraints. The 
following reductions have straightforward proofs.

\setCoqFilename{red_undec}
\begin{fact}
	\label{prop:embed_sig}
        One has reductions for the three statements below (for $n\ge 2$):
        \begin{enumerate}
        \coqitem[FSAT_REL_2ton] $\FSAT(\Void; \{P^2\})\red \FSAT(\Void; \{P^n\})$;
	\coqitem[FSAT_RELn_ANY] $\FSAT(\Void; \{P^n\})\red \FSAT(\Sigma)$ if\/ $\Sigma$ contains an\/ $n$-ary relation symbol;
        \coqitem[FSAT_FUNnREL1_ANY] $\FSAT(\{f^n\};\{Q^1\})\red \FSAT(\Sigma)$ if\/ $\Sigma$ contains an\/ $n$-ary fun.\ and a unary rel.
        \end{enumerate}
\end{fact}


\section{Decidability Results}

Complementing the previously studied negative results, we now examine the conditions allowing for decidable satisfiability problems.

\label{sec:decidability}

\setCoqFilename{fo_sat_dec}
\begin{lemma}[FSAT over a fixed domain][FSAT_in_dec]
\label{lem:FSAT_in}
Given a discrete signature $\Sigma$ and a discrete and
finite type $D$, one can decide whether or not a formula
over $\Sigma$ has a (finite) model over domain $D$.
\end{lemma}

\begin{proofqed}
By Fact~\ref{fact:satisfaction_dec},
satisfaction in a given finite model
is decidable. 
It is also \coqlink[FO_model_equiv_spec]{invariant under extensional equivalence}, so we only need to show that there are 
\coqlink[finite_t_model_upto]{finitely many (decidable) models}
over $D$ up to extensional equivalence\rlap.%
\footnote{Without discreteness of $\Sigma$, it is impossible
to build the list of models over $D=\Bool$.}
\end{proofqed}

\begin{lemma}[][fo_form_fin_discr_dec_SAT_pos]
\label{lem:FSAT_in_pos}
A formula over a signature\/ $\Sigma$
has a finite and discrete model if and only if it has a (finite) model over\/
$\Fin n$ for some\/ $n:\Nat$.
\end{lemma}

\begin{proofqed}
If $\phi$ has a model over
a discrete and finite domain $D$, by Corollary~\ref{coro:fin_type}, 
one can bijectively map $D$ to $\Fin n$  and transport
the model along this bijection.
\end{proofqed}

\setCoqFilename{red_dec}
\begin{lemma}[][FSAT_MONADIC_DEC]
\label{lem:monadic_fol}
$\FSAT(\Void;\Preds)$ is decidable if\/ $\Preds$ is discrete with uniform arity\/ $1$.
\end{lemma}

\begin{proofqed}
By Lemma~\ref{lem:signature_finite}, we can assume $\Preds=\Fin n$ w.l.o.g.\enspace
We show that if $\phi$ has a finite model then it must have a 
model over domain $\{\vec v:\Bool^n\to\Bool\mid b\,\vec v = \btrue\}$ for
some Boolean subset $b:(\Bool^n\to\Bool)\to\Bool$.
Up to extensional equivalence, there are only finitely many such subsets 
$b$ and we conclude with Lemma~\ref{lem:FSAT_in}.
\end{proofqed}

\begin{lemma}[][FSAT_MONADIC_11_FSAT_MONADIC_1]
\label{lemma:remove_funcs}
For any finite type\/ $\Preds$ of relation
symbols and signatures of \emph{uniform arity\/ $1$}, we have a reduction
$\FSAT(\Fin n;\Preds)\red\FSAT(\Void;\List\,{\Fin n}\times\Preds+\Preds)$.
\end{lemma}

\begin{proofqed}
We implemented a proof somewhat inspired by that of Proposition~6.2.7 (Grädel) 
in~\cite[pp.~251]{borger1997classical} 
but the invariant suggested in the iterative process described there did not work 
out formally
and we had to proceed in a single conversion step instead, switching
from single symbols to lists of symbols.
\end{proofqed}

If functions or relations have arity $0$, one can always
lift them to arity $1$ using a fresh variable (of arbitrary value),
like in~Fact~\ref{prop:three_reds}, item~(1).

\begin{fact}[][FSAT_FULL_MONADIC_FSAT_11]
\label{prop:rem_cst_props}
The reduction\/ $\FSAT(\Funcs;\Preds)\red\FSAT(\Funcs^1;\Preds^1)$ holds when
all arities in\/ $\Sigma$ are at most 1, 
where\/ $\Funcs^1$ and\/ $\Preds^1$ denote arities uniformly updated to $1$.
\end{fact}

\section{Signature Classification}
\label{sec:trakh_full}

We conclude with the exact classification of $\FSAT$ regarding enumerability, decidability, and undecidability depending on the properties of the signature. 

\setCoqFilename{red_enum}
\begin{theorem}[][FSAT_opt_enum_t]
\label{thm:FSAT_enum}
Given\/ $\Sigma=(\Funcs;\Preds)$ where both\/ $\Funcs$
and\/ $\Preds$ are data types, the finite satisfiability problem
for formulas over\/ $\Sigma$ is  enumerable.
\end{theorem}

\begin{proofqed}
Using Theorem~\ref{thm:quotient_FSAT} and Lemmas~\ref{lem:FSAT_in} and~\ref{lem:FSAT_in_pos}, one constructs 
a predicate $Q:\Nat\to \Formula_\Sigma\to\Bool$ s.t.\ 
\coqlink[FSAT_rec_enum_t]{$\FSAT(\Sigma)\,\phi\toot\exists n.\,Q\,n\,\phi = \btrue$}. 
Then, it is easy to build a \coqlink[FSAT_opt_enum_t]{computable enumeration} $e:\Nat\to\Opt\,\Formula_\Sigma$ of $\FSAT(\Sigma):\Formula_\Sigma\to\Prop$.
\end{proofqed}

\setCoqFilename{red_dec}
\begin{theorem}[Full Monadic FOL][FULL_MONADIC]
	\label{thm:full_monadic_fol}
	$\FSAT(\Sigma)$ is decidable if\/ $\Sigma$ is discrete with arities less or equal than\/ $1$, or
	if all relation symbols have arity\/ $0$.
\end{theorem}

\begin{proofqed}
	If all arities are at most $1$, then by Fact~\ref{prop:rem_cst_props}, we can assume $\Sigma$ of uniform arity $1$. 
	Therefore, for a formula $\phi$ over $\Sigma$ with uniform arity $1$, we
	need to decide $\FSAT$ for $\phi$. By Theorem~\ref{lem:signature_finite}, we can compute a 
	signature $\Sigma_{n,m}=(\Fin n;\Fin m)$ and a formula $\psi$ over $\Sigma_{n,m}$
	equi-satisfiable with $\phi$.
	Using the reduction of Lemma~\ref{lemma:remove_funcs}, we compute a formula $\gamma$, 
	equi-satisfiable with $\psi$, over a discrete signature of uniform arity $1$, void of functions. 
	We decide the satisfiability of $\gamma$ by Lemma~\ref{lem:monadic_fol}.
	
	If all relation symbols have arity $0$,
	regardless of $\Funcs$, 
	no term can occur in formulas,
	hence neither can function symbols. Starting from $\phi$ over $\Sigma=(\Funcs;\Preds^0)$
	where only $\Preds$ is assumed discrete, we compute an equi-satisfiable formula $\psi$ over $\Sigma'=(\Void;\Preds^0)$
	and we are back to the previous case.
\end{proofqed}

\setCoqFilename{red_undec}
\begin{theorem}[Full Trakhtenbrot][FULL_TRAKHTENBROT]
\label{thm:full_trakhtenbrot}
        If\/ $\Sigma$ contains either an at least binary relation symbol or
        a unary relation symbol together with an at least binary function symbol, then\/
        $\BPCP$ reduces to $\FSAT(\Sigma)$.
\end{theorem}

\begin{proofqed}
	By Theorems~\ref{thm:BPCP_FSATEQ}, \ref{thm:uninterpret} and~\ref{thm:discrete_to_binary},
        Lemma~\ref{lem:sig2_sig21}, and Fact~\ref{prop:embed_sig}.
\end{proofqed}

\begin{corollary}
	For an enumerable and discrete signature $\Sigma$ furthermore satisfying the conditions in Theorem~\ref{thm:full_trakhtenbrot},\/ $\FSAT(\Sigma)$ 
        is both enumerable and undecidable, thus, more specifically, not co-enumerable.
\end{corollary}

\begin{proofqed}
	Follows by Facts~\ref{fact:reduction} and \ref{fact:BPCP_undec}.
\end{proofqed}

Notice that even if the conditions on arities of Theorems~\ref{thm:full_monadic_fol}
and~\ref{thm:full_trakhtenbrot}  fully classify discrete signatures, it is not
possible to decide which case holds unless the signature is furthermore finite.
For a given formula $\phi$ though, it is always possible to render it in the finite signature of used symbols.

\section{Discussion}
\label{sec:discussion}


The main part of our Coq development directly concerned with the 
\href{https://github.com/uds-psl/coq-library-undecidability/tree/trakhtenbrot_ijcar/theories/TRAKHTENBROT}{classification of finite satisfiability} 
consists of 10k loc, in addition to 3k loc of (partly reused)
\href{https://github.com/uds-psl/coq-library-undecidability/tree/trakhtenbrot_ijcar/theories/Shared/Libs/DLW}{utility libraries}.
Most of the code comprises the signature transformations
with more than 4k~loc for reducing discrete signatures to membership.
Comparatively, the initial reduction from $\BPCP$ to $\FSATEQ(\SBPCP)$ takes less than 500~loc.

Our mechanisation of first-order logic in principle follows previous developments~\cite{ForsterCPP,ForsterEtAl:2019:Completeness} but also differs in a few aspects.
Notably, we had to separate function from relation signatures to be able to express distinct signatures that agree on one sort of symbols computationally.
Moreover, we found it favourable to abstract over the logical connectives in form of $\binop$ and $\quant$ to shorten purely structural definitions and proofs.
Finally, we did not use the Autosubst 2~\cite{AutoSubst2} support for de Bruijn syntax to avoid its current dependency on the functional extensionality axiom.

We refrained from additional axioms since we included our development in the growing \href{https://github.com/uds-psl/coq-library-undecidability}{Coq library of synthetic 
undecidability proofs}~\cite{library_coqpl}.
In this context, we plan to generalise some of the intermediate signature reductions so that they become reusable for 
other undecidability proofs concerning first-order logic over arbitrary models.

As further future directions, we want to explore and mechanise the direct consequences of Trakhtenbrot's theorem such as 
the undecidability of query containment and equivalence in data base theory
or the undecidability of separation logic~\cite{BROCHENIN2012106,10.1007/3-540-45294-X_10}.
Also possible, though rather ambitious, would be to mechanise the classification of first-order satisfiability with regards to the quantifier prefix as
comprehensively developed in~\cite{borger1997classical}.
Finally, we plan to mechanise the undecidability of semantic entailment and syntactic deduction in first-order axiom systems such as ZF set theory and Peano arithmetic.


\bibliographystyle{splncs04}
\bibliography{paper}

\if\isijcar1
	\end{document}
\fi

\appendix

\section{Tools for Finite Types}

\label{appendix:tools_finite}


\setBaseUrl{{http://www.ps.uni-saarland.de/extras/fol-trakh/website/Undecidability.Shared.Libs.DLW.Utils.}}
\setCoqFilename{php}
\begin{theorem}[Finite PHP]
\label{appendix:thm:finite_php}
Let\/ $R:X\to Y\to\Prop$ be a binary relation 
and\/ $l:\List\,X$ and $m:\List\,Y$ be two lists where\/
$m$ is shorter than\/ $l$ $(\clen m < \clen l)$. 
If\/ $R$ is total from\/ $l$ to\/ $m$ 
$(\forall x.\, x\inl l \to \exists y.\, y\inl m \land R\,x\,y)$ then
the values at two distinct positions in\/ $l$ are related to the same\/ $y$ in\/ $m$,
i.e.\/ there exist\/ $x_1,x_2\inl l$ and $y\inl m$ such that\/ $l$ has shape\/ 
$l=\cdots\capp x_1\ccons \cdots\capp x_2\ccons \cdots$ and\/ $R\,x_1\,y$ and\/ $R\,x_2\,y$.
\end{theorem}

\begin{proofqed}
We start with the case where $R$ is the identity relation $=_X$ on $X$, hence
we want to establish that $l$ contains a duplicate. 
 We first prove the following \coqlink[length_le_and_incl_implies_dup_or_perm]{generalised statement}:
if $\clen m\le\clen l$ and $l\incl m$ (i.e.\ $\forall x.\, x\inl l \to x\inl m$)
then either $l$ contains a duplicate 
or $l$ and $m$ are permutable. We establish the generalised statement
by structural induction on $m$.

In particular, when $\clen m<\clen l$ then $l$ and $m$ cannot be permutable
(because permutations preserve length), hence \coqlink[finite_pigeon_hole]{$l$ must contain a duplicate}.
Generalizing from $=_X$ to an arbitrary relation $R:X\to Y\to\Prop$ is then a simple
exercise.
\end{proofqed}

\setBaseUrl{{http://www.ps.uni-saarland.de/extras/fol-trakh/website/Undecidability.}}
\setCoqFilename{Shared.Libs.DLW.Wf.wf_finite}
\begin{fact}[]
	\label{appendix:fact:wf}
	Every strict order on a finite type is \coqlink[wf_strict_order_finite]{well-founded}.
\end{fact}

\begin{proofqed}
	For a constructive proof, one can for instance show that descending chains cannot contain a duplicate
        (otherwise this would give an impossible cycle in a strict order), hence by the PHP,
        the length of descending chains is bounded by the length of the enumerating list of the finite type.
\end{proofqed}

\setCoqFilename{Shared.Libs.DLW.Utils.fin_quotient}
\begin{theorem}[Finite decidable quotient]
Let\/ ${\sim}:X\to X\to\Prop$ be a decidable equivalence with
$\SigType{l_r:\List\,X}{\forall x\exists y.\,y\inl l_r\land x\sim y}$, i.e.\/
finitely many equivalence classes.\footnote{Hence $l_r$ denotes a list of \emph{representatives} of equivalence classes.}\enspace
Then one can compute the \coqlink[decidable_EQUIV_fin_quotient]{quotient}\/ $X/{\sim}$ onto\/ $\Fin n$ for some $n$, i.e.\/
$n:\Nat$,\/ $c:X\to\Fin n$ and\/ $r:\Fin n\to X$ s.t.\/
$\forall p.\,c\,(r\,p) = p$ and\/ $\forall x \,y.\,x\sim y\toot c\,x = c\, y$.
\end{theorem}

\begin{proofqed}
From the list $l_r$ of representatives of equivalence classes, remove duplicate representatives
using the strong decidability of $\sim$. This gives a list $l'_r$ which now contains exactly 
one representative for each equivalence class. Convert $l'_r$ to a vector $\vec v$. The function $r$ 
(representative) is defined by $r\cdef \lambda p.\,\vec v_p$. The function $c$ (for class) is
simple search: $c\,x$ is the first (and unique) $p$ such that $\vec v_p \sim x$.  
\end{proofqed}

\section{Discrete Domains}

\label{appendix:discrete}


We give an account of the proof of Theorem~\ref{thm:fo_indist} 
(numbered~\ref{appendix:thm:fo_indist} later in Appendix~\ref{appendix:discrete}) 
stating the first-order indistinguishability $\foindist$ up to 
two given lists $\lf $ and $\lp $ of function and relation symbols
respectivelly, and defined by 
$$x\foindist y~\cdef~\forall (\rho:\Nat\to D)\,\phi.\,\funcs\phi\subseteq\lf\to
\preds\phi\subseteq\lp\to\MM\vDash_{x\cdot\rho}\phi\toot \MM\vDash_{y\cdot\rho}\phi$$
is a strongly decidable equivalence and a congruence for all the symbols in $\lf/\lp $.
To remain simple, we avoid displaying the dependency on $\lf $, $\lp $, $D$ and
$\MM$ in the notation $\foindist$ as they remain fixed in this section anyway.
Equivalence and congruence of $\foindist$ are easy, but congruence
is of course limited to $\lf /\lp $:
$$\begin{array}{c}
\forall (f:\Funcs)\,(\vec v\, \vec w:D^{\arity f}).\, f\inl \lf \to (\forall i:\Fin{\arity f}.\, \vec v_i\foindist \vec w_i)
\to f^\MM\,\vec v\foindist f^\MM\,\vec w\\
\forall (P:\Preds)\,(\vec v\, \vec w:D^{\arity P}).\, P\inl \lp \to (\forall i:\Fin{\arity P}.\, \vec v_i\foindist \vec w_i)
\to P^\MM\,\vec v\toot P^\MM\,\vec w\\
\end{array}$$
However, the definition of $\foindist$ hints at no clue for its decidability. 
We therefore switch to an alternate definition of $\foindist$
as a bisimulation\rlap.\footnote{That is the greatest fixpoint of an $\omega$-continuous operator.}\enspace 
Using Kleene's fixpoint theorem, we would get ${\foindist}$ as $\bigcap_{n<\omega}\Fb^n(\lambda uv.\top)$ 
for some $\omega$-continuous operator~$\Fb$.
Hopefully, finiteness would ensure that only finitely many (as opposed to $\omega$) 
iterations of the operator $\Fb$ are needed for the fixpoint to be 
reached, hence preserving finitary properties such as decidability. 

\setBaseUrl{{http://www.ps.uni-saarland.de/extras/fol-trakh/website/Undecidability.TRAKHTENBROT.}}
\setCoqFilename{discrete}

\smallskip

So let us define the operators
$\coqlink[fom_op1]{\Ff},\coqlink[fom_op2]{\Fp}:(D\to D\to\Prop)\to(D\to D\to\Prop)$ that
map a binary relation ${\genrel}:D\to D\to\Prop$ to
$$\begin{array}{r@{~\cdef~}l}
\Ff({\genrel})\,x\,y & \forall f.\, f\inl \lf \to \forall(\vec v:D^{\arity f})\,(i:\Fin{\arity f}).\,
   \genrel\,\bigl(f^\MM\,\subst{\vec v}xi\bigr)\,\bigl(f^\MM\,\subst{\vec v}yi\bigr)\\
\Fp({\genrel})\,x\,y & \forall P.\, P\inl \lp \to \forall(\vec v:D^{\arity P})\,(i:\Fin{\arity P}), 
   P^\MM\,\subst{\vec v}xi \toot P^\MM\,\subst{\vec v}yi.\\
\end{array}$$

\begin{fact}
\label{appendix:fact:bisim}
The following results hold for the operator $\Ff$ (resp.\ $\Fp$).
\begin{enumerate}
\coqitem[fom_op_mono] $\Ff$ is monotonic, i.e.\/ ${\genrel} \subseteq{\genrel'}\to \Ff({\genrel})\subseteq \Ff({\genrel'})$;
\coqitem[fom_op_continuous] $\Ff$ is continuous, i.e.\/ $\bigcap_{n} \Ff({\genrel_n}) \subseteq \Ff\bigl(\bigcap_{n}{\genrel_n}\bigr)$ with decreasing\/ 
$(\genrel_n)_{n<\omega}$;
\coqitem[fom_op_id] $\Ff$ preserves reflexivity, i.e.\/ ${=_D}\subseteq \Ff({=_D})$;
\coqitem[fom_op_sym] $\Ff$ preserves symmetry, i.e.\/ $\Ff^{-1}({\genrel})\subseteq \Ff({\genrel^{-1}})$;
\coqitem[fom_op_trans] $\Ff$ preserves transitivity, i.e.\/ $\Ff({\genrel})\circ \Ff({\genrel})\subseteq \Ff({\genrel}\circ{\genrel})$;
\coqitem[fom_op_dec] $\Ff$ preserves decidability, i.e.\/ if\/ ${\genrel}$ is decidable then so is\/ $\Ff({\genrel})$. 
\end{enumerate}
Hence the combination\/ $\coqlink[fom_op]{\Fb({\genrel})}\cdef \Ff({\genrel})\cap \Fp({\genrel})$ also preserves these properties.
\end{fact}

\begin{proofqed}
The proofs of items (1)-(5) are easy, even without assuming boundedness by $\lf /\lp $. However,
to ensure  the preservation of decidability (6), that bound is essential for the quantification
over $\lf$ (resp.\ $\lp$) to stay finite. Notice that since $D$ is finite then so is $D^{\arity f}$
and the remaining quantifications over
$\vec v:D^{\arity f}$ and $i:\Fin{\arity f}$ are finite quantifications again. Hence, 
they behave as finitary conjunctions and thus preserve decidability. Notice that compared
to $\Ff$, the case of $\Fp$ is degenerated because it is constant w.r.t.\ $\genrel$.
\end{proofqed}

\begin{theorem}[][fom_eq_fol_characterization]
First-order indistinguishability\/ $\foindist$ up to\/ $\lf /\lp $ is extensionally equivalent 
to\/ $\bisim$ (Kleene's greatest fixpoint of\/ $\Fb$), i.e.\ for any\/ $x,y:D$ we have 
$$x\foindist y~\toot~x\bisim y\quad\text{where}\quad
x\bisim y~\cdef~\forall n:\Nat.\,\Fb^n(\lambda uv.\top)\,x\,y.$$
\end{theorem}

\begin{proofqed}
For the $\to$ implication, it is enough to show that $\foindist$ is
a pre-fixpoint of $\Fb$, i.e.\ ${\foindist}\subseteq \Fb({\foindist})$, and
we get this result using \coqlink[fo_bisimilar_fom_eq]{suitable substitutions}. The converse implication
$\leftarrow$ follows from the fact that $\bisim$ is a fixpoint of $\Fb$,
hence it is a congruence for every symbol in\/ $\lf /\lp $, so
$x\bisim y$ entails that \coqlink[fom_eq_fo_bisimilar]{formulas built from $\lf /\lp $ cannot 
distinguish $x$ from $y$}.
\end{proofqed}

With $\bisim$ we have a more workable characterization of $\foindist$ but still no decidability
result for it since the quantification over $n$ in 
$\forall n:\Nat.\, \Fb^n(\lambda uv.\top)\,x\,y$ ranges over the infinite domain $\Nat$.
We now establish that the greatest fixpoint is reached after finitely
many iterations of $\Fb$. Classically one would argue that $\Fb$ operates over
the finite domain of binary relations over $D$ and since the sequence
$\lambda n.\,\Fb^n(\lambda uv.\top)$ cannot decrease strictly forever (by the PHP), 
it must stay constant after at most $n_0\cdef 2^{d\times d}$ 
iterations where $d\cdef\mathrm{card}\,D$.

Unfortunately, no such reasoning is constructively possible since even for the unit
type $\Unit$, there is no list enumerating the predicates $\Unit\to\Prop$. However,
there is a notion of weak powerset. Recall that a predicate $p:X\to\Prop$ is
\emph{weakly decidable} if it satisfies $\forall x.\, p\,x\lor\lnot p\,x$.

\setBaseUrl{{http://www.ps.uni-saarland.de/extras/fol-trakh/website/Undecidability.Shared.Libs.DLW.Utils.}}
\setCoqFilename{fin_upto}
\begin{lemma}[Weak powerset][finite_t_weak_dec_powerset]
\label{appendix:lem:wposet}
For every finite type\/ $X$, one can compute a list\/ $ll:\List(X\to\Prop)$ which contains every
weakly decidable predicate in\/ $X\to\Prop$ up to extensionality, i.e.\
$\forall p:X\to\Prop.\,(\forall x.\, p\,x\lor\lnot p\,x)\to\exists q.\, q\inl ll\land \forall x.\, p\,x\toot q\,x$.  
\end{lemma}

\begin{proofqed}
The list $ll$ is built by induction on the list $l_X:\List\,X$ enumerating $X$. 
If $l_X$ is $\cnil$ then $X$ is a void type and thus $ll \cdef (\lambda z.\top)\ccons\cnil$ fits.
If $l_X$ is $x\ccons l$ then we apply the induction hypothesis to $l$ and
get $ll$ for the finite sub-type composed of the elements of $l$ and
we define $ll'\cdef (\lambda p\,z.\, x\neq z\land p\,z)\map ll\capp 
                     (\lambda p\,z.\, x= z\lor p\,z)\map ll$. We
check that $ll'$ contains every weakly decidable predicate over $x\ccons l$.
Notice that $\clen{ll'}=2\clen{ll}$ in the induction step, hence one could easily 
show that $\clen{ll}=2^{\clen{l_X}}$, recovering the cardinality of the
(classical) powerset.
\end{proofqed}

Notice that the weak powerset contains all weakly decidable predicates but 
not every predicate in it is necessarily weakly decidable\rlap.\footnote{Unless $X$ is
moreover discrete.}\enspace
Now we show that $\lambda n.\, \Fb^n(\lambda uv.\top)$
converges after finitely many steps.

\setBaseUrl{{http://www.ps.uni-saarland.de/extras/fol-trakh/website/Undecidability.TRAKHTENBROT.}}
\setCoqFilename{discrete}
\begin{theorem}[][fom_eq_finite]
\label{appendix:thm:bisim_finite}
One can compute\/ $n:\Nat$ such that\/ ${\bisim}$ is equivalent to\/ $\Fb^n(\lambda uv.\top)$.
\end{theorem}

\begin{proofqed}
By a \coqlink[finite_t_weak_dec_rels]{variant} of Lemma~\ref{appendix:lem:wposet}, 
we compute the weak powerset of $D\to D\to\Prop$,\footnote{Via $D\to D\to\Prop\simeq D\times D\to\Prop$,
and finiteness of $D\times D$.} i.e.\
a list $ll$ containing every weakly decidable binary relation over $D$, up to extensional equivalence. 
Since $\lambda uv.\top:D\to D\to\Prop$ is strongly decidable
and $\Fb$ preserves (both weak and) strong decidability, the sequence
$\lambda n.\, \Fb^n(\lambda uv.\top)$ is contained in the list $ll$, up to extensionality.
Hence by Theorem~\ref{thm:finite_php} (PHP)\rlap,\footnote{And here we really need a finite PHP 
over \emph{non-discrete types}.} after $\clen{ll}$ steps, there must have
been a duplicate, i.e.\ there exists $a<b\le \clen{ll}$ such that $\Fb^a(\lambda uv.\,\top)$
and $\Fb^b(\lambda uv.\top)$ are extensionally equivalent. However the values of $a$ and $b$
are not computed by the PHP but we can still deduce that $\Fb^n(\lambda uv.\top)$
must be stalled after $n=a$, hence a fortiori after $n=\clen{ll}$. 
Hence $\Fb^{\clen{ll}}(\lambda uv.\top)$ 
is extensionally equivalent to $\bisim$. 
\end{proofqed}

\setCoqFilename{discrete}
\begin{theorem}[]
\label{appendix:thm:fo_indist}
First-order indistinguishability up to\/ $\lf /\lp $ is a strongly decidable
equivalence and a congruence for all the symbols in\/ $\lf /\lp $. 
\end{theorem}

\begin{proofqed}
Remember that the real difficulty was strong decidability.
By Theorem~\ref{appendix:thm:bisim_finite}, the operator $\Fb$ reaches its fixpoint
$\bisim$ after finitely many steps, and by Fact~\ref{appendix:fact:bisim} item~6, $\Fb$
preserves decidability, hence by an obvious induction, $\bisim$ is decidable. 
By Theorem~\ref{coq:fom_eq_fol_characterization}, the equivalent indistinguishability
relation $\foindist$ is decidable.
\end{proofqed}

As a side remark, notice that we also show that $\Fb$ preserves first-order definability.
A relation $\genrel:D\to D\to\Prop$ is \emph{first-order definable} if there
is a formula $\phi$ built only from $\lf /\lp $ such that
$\forall\rho.\, \genrel\,(\rho\,x_0)\,(\rho\,x_1)\toot  \MM\vDash_\rho \phi$.
By Theorem~\ref{appendix:thm:bisim_finite}, there is thus a \coqlink[fom_eq_form_sem]{first-order
formula which is equivalent to first-order indistinguishability in $\MM$}. 
Since its semantics does not depend on variables other that $x_0$ and $x_1$,
one can remap all other variables to e.g.\ $x_0$ hence we can
even ensure that the first-order formula defining $\foindist$ contains
only two free variables, namely $x_0$ and $x_1$.

\end{document}